\documentclass[12pt]{article}

\usepackage{scicite}
\usepackage{graphicx}
\usepackage{amssymb}
\usepackage{amsmath}

\usepackage{times}

\newcommand{\unit}[1]{\ensuremath{\, \mathrm{#1}}}

\newenvironment{sciabstract}{%
\begin{quote} \bf}
{\end{quote}}

\newcounter{lastnote}

\title{Real-time detection of an extreme scattering event: constraints on Galactic plasma lenses}

\author
{Keith W. Bannister,$^{1\ast}$ Jamie Stevens,$^{1}$ Artem V. Tuntsov,$^{2}$ Mark A. Walker, $^{2}$ \\
Simon Johnston, $^{1}$ Cormac Reynolds, $^{3}$ Hayley Bignall $^{3}$\\
\\
\normalsize{$^{1}$CSIRO Astronomy and Space Science, PO Box 76, Epping NSW 1710, Australia}\\
\normalsize{$^{2}$Manly Astrophysics, 3/22 Cliff St, Manly 2095, Australia}\\
\normalsize{$^{3}$International Centre for Radio Astronomy Research - Curtin University, Perth, Australia}\\
\\
\normalsize{$^\ast$To whom correspondence should be addressed; E-mail:  keith.bannister@csiro.au.}
}

\date{}

\begin{document}

\baselineskip24pt

\maketitle

\begin{sciabstract}
Extreme scattering events (ESEs) are distinctive fluctuations in the brightness of astronomical radio sources caused by occulting plasma lenses in the interstellar medium. The inferred plasma pressures of the lenses are  $\sim 10^3$ times the ambient pressure, challenging our understanding of gas conditions in the Milky Way. Using a new survey technique, we have discovered an ESE while it was in progress. We report radio and optical follow-up observations. Modelling of the radio data demonstrates that the lensing structure is a density enhancement and that the lens is diverging, ruling out one of two competing physical models. Our technique will uncover many more ESEs, addressing a long-standing mystery of the small-scale gas structure of the Galaxy.
\end{sciabstract}

Distinctive variations in the radio light curves of quasars were first identified serendipitously while using the Green Bank Interferometer in support of routine astrometric observations \cite{Fiedler87ese}, and were named `extreme scattering events' (ESEs). It was immediately recognised that the cause could not be intrinsic to the quasar, but rather must be a foreground propagation phenomenon. The presumed model was occulting clouds at a distance $\sim 1$~kpc with size $\sim 10^{14}$~cm, and high free electron density $\sim 10^{3}$~cm$^{-3}$. One immediate problem posed by these observations was that the combination of high density and the fact that the gas must be ionised implies pressures that are three orders of magnitude larger than the typical pressure in the diffuse interstellar medium (ISM) \cite{Kulkarni88}. While such pressures have been observed in a very small fraction of interstellar gas \cite{Jenkins01, Jenkins11}, the fact that ESEs are relatively common raises the question of how such structures can form and live long enough to be detected as ESEs. There is general agreement \cite{Romani87, Rickett97} (with some exceptions  \cite{Hamidouche07}) that Kolmogorov turbulence in the ISM cannot generate such large, discrete and long-lived over-pressured regions, so two competing physical models have been proposed: plasma sheets seen edge-on \cite{Romani87, Pen12} provide the required column density variation, but yield light curves which are at odds with the data \cite{Walker07}, while models of cold self-gravitating clouds \cite{Henriksen95, Walker98, Walker07} solve the pressure problem but in turn imply that the clouds must make up a substantial fraction of the mass of the Galaxy.

The interpretation of ESEs remains uncertain because existing narrow-band radio monitoring has only limited use in determining the lens properties, and because no ESEs have been detected at any other wavelength or modality while they were in progress. To date, only one ESE has been followed-up while it was in progress: Q1741$-$038 showed no changes in H\,{\sc i} absorption \cite{Lazio01hi}, or rotation measures \cite{Clegg96}, while VLBI \cite{Lazio00} showed a marginal increase in the observed size of the source. To date, no observations of ESEs have been made at optical, infrared or X-ray wavelengths.

Historically, ESEs have been difficult to find, and required large investments in telescope time for only limited number of detections. \cite{Lazio01lc}, building on the work of \cite{Fiedler94}, searched the largest available dataset for ESEs. Their data contained dual-frequency observations of 149 sources every $\sim $2 days spanning 17 years. With a total of $\sim $~1200~source-years of observations, they had discovered only $\sim 15$ events. Only a handful of other events have been discovered serendipitously \cite{Cognard93, Cimo02, Maitia03, Savolainen08, Senkbeil08, Pushkarev13}, but in no cases were follow-up observations made that could shed light on the physical nature of the lensing structures.

Real-time detection of ESEs is the key to their understanding, as most properties of the lens are only measurable with intensive follow-up observations while the ESE is in progress. For example, well-sampled radio light curves permit the the electron column density profile to be determined. Very Long Baseline Interferometry (VLBI) can measure the angular scale and geometry of the lens in the plane of the sky. Temporary reddening of the background source at optical wavelengths can reveal the presence of dust in the lens, while absorption lines against the background source can measure the composition and physical conditions in the lens. Finally, changes in rotation measure can probe the lens' magnetic field. A measurement of any of these properties would represent a breakthrough, and would help solve the puzzle of the origin of these lenses, and how they form and survive. Real-time detection is required to deploy expensive follow-up resources to maximum effect.

We have developed an efficient technique for finding ESEs in real-time, by exploiting the wideband receiver and correlator on the Australia Telescope Compact Array (ATCA) \cite{Wilson11}. The key to the technique is that when a plasma lensing event is in progress, the radio spectrum of the background source changes from a featureless continuum to one which is highly structured. This is a result of the $\lambda^2$ dependence of the plasma refractive index \cite{Stix92}, which moves the lens focus closer to the observer at some frequencies than at others. We selected a sample \cite{SciSupplementary} of $\sim 1000$ active galactic nuclei (AGN) and observed this sample once per month, obtaining spectra in the range 4-8~GHz. With only 50~s on target we obtain a spectrum with signal-to-noise $> 50$ in a  64~MHz channel, which is sufficient to identify an ESE in progress, simply by searching for a spectrum which is not well modelled by a smooth power law, and which has changed significantly between epochs. Each survey run requires 24~hr observing time to obtain spectra for all targets. Instantaneous identification is the key requirement for follow-up studies, which yield more useful data when initiated early in the event. The false detection rate is low, as the ESE signature is much more strongly frequency-dependent than intrinsic AGN variability, which is generally modelled as broad-band synchrotron emission \cite{Valtaoja92}.

\begin{figure}
\centering
\includegraphics[width=0.8\textheight]{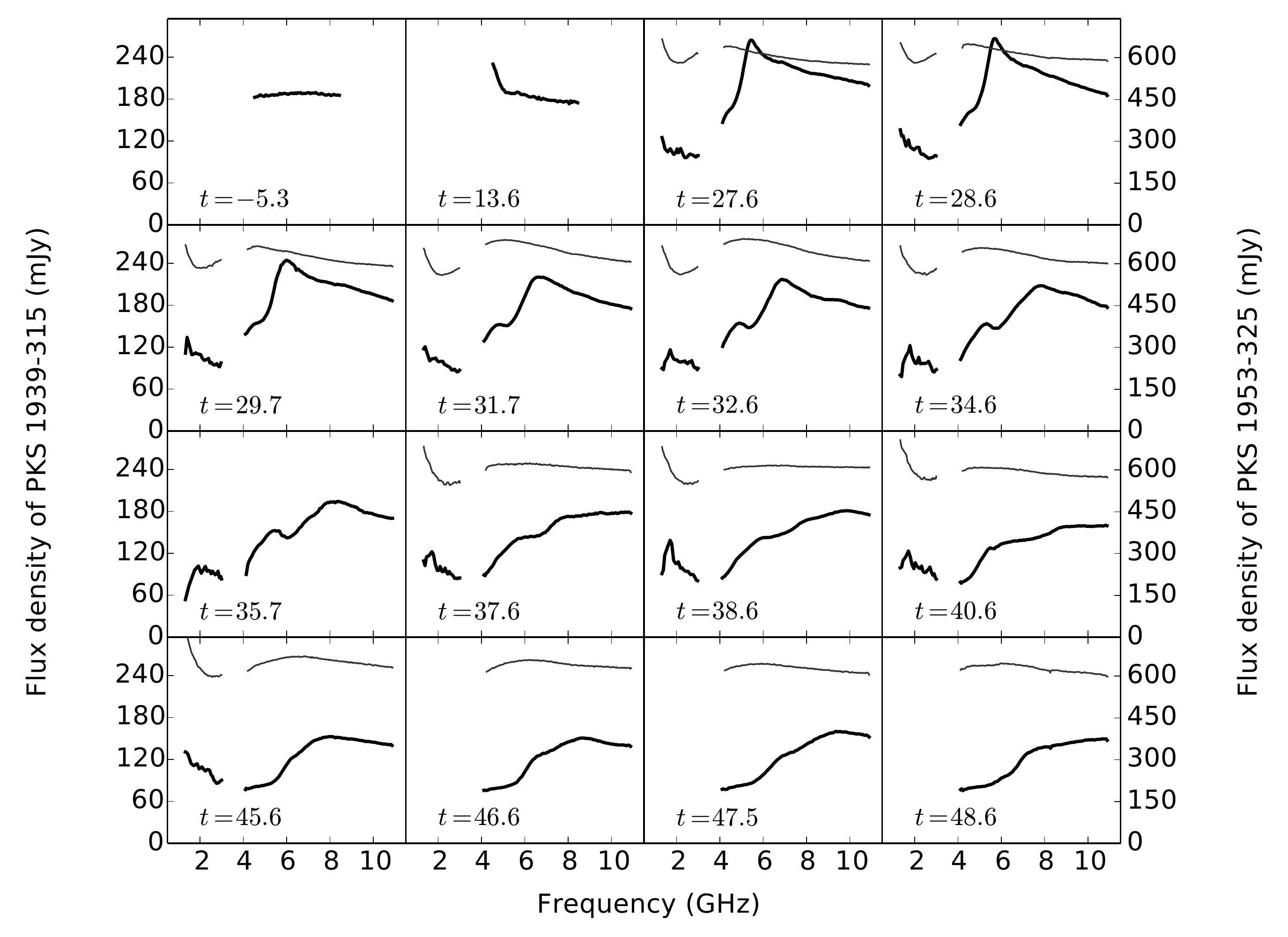}
\caption{\label{fig:atca_stack} {\bf The first 16 ATCA radio spectra of PKS~1939$-$315 (thick black line), and near-simultaneous spectra of ATCA calibrator PKS~1953$-$325 (thin gray line)}. $t=\rm{MJD}-56800$ is the observing day. The first survey spectrum at t=-5.3 showed no evidence of an ESE in PKS~1939$-$315, while the discovery spectrum during the survey on t=13.6 showed strong evidence, with a sharp up-tick below 5~GHz. On t=27.6 we started wide-band, regular monitoring of PKS~1939$-$315 and an ATCA calibrator PKS~1953$-$325. Where both sources were observed, they were observed within 20~min of each other. PKS~1953$-$325 was not observed on the first two epochs, and is weakly variable in the monitoring.  All spectra have been averaged to 64~MHz channels. Measurement errors from thermal noise in each spectrum are 0.5~mJy RMS, less than the thickness of the lines. ATCA receivers do not cover the frequencies 3.1 to 3.9~GHz.}
\end{figure}

\begin{figure}
\centering
\includegraphics[height=0.8\textheight]{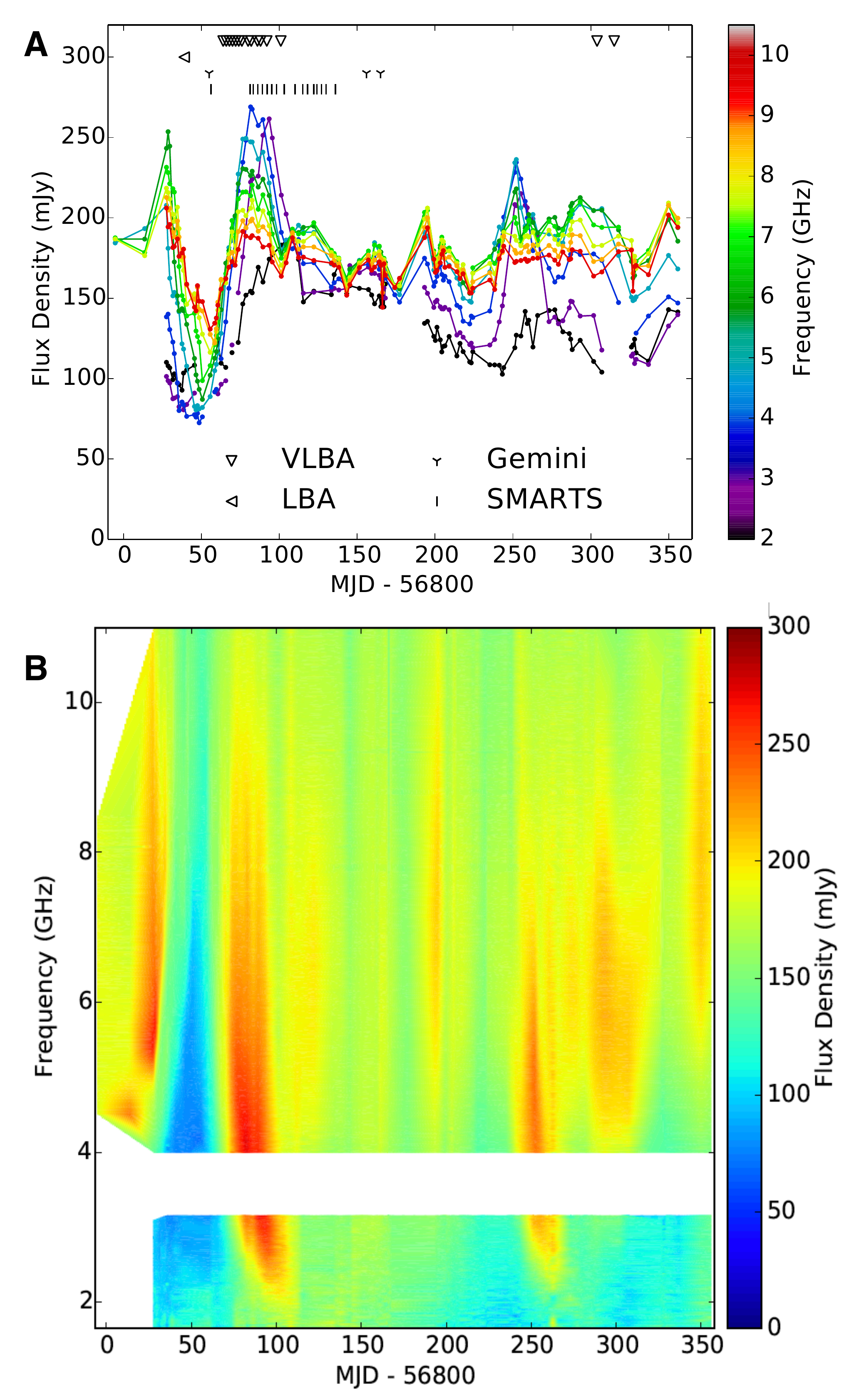}
\caption{\label{fig:atca_ds} {\bf 104 epochs of ATCA radio data of PKS~1939$-$315.} Panel A: Multi-frequency light curve comprising nine 64~MHz channels centred every 1~GHz from 2 to 10~GHz. Thermal noise at each point is 0.5~mJy, less than the thickness of the lines. The symbols above the light curve indicate the days when follow-up observations were obtained using the VLBA, the Gemini 8~m, and SMARTS 1.3~m. Panel B: the dynamic spectrum averaged to 4~MHz frequency resolution, sampled on a 1 day grid, and linearly interpolated between observing epochs. Thermal noise at each spectral point is 2~mJy. ATCA receivers do not cover the frequencies 3.1 to 3.9~GHz.}
\end{figure}

We discovered an ESE toward PKS~1939$-$315 on 2014 June 05  (Modified Julian Date (MJD) 56813),  two months after beginning our survey, when it showed a substantial change in spectral shape near 4.5~GHz. A power-law fit to this discovery spectrum has a $\chi^2/N_{d.o.f} = 12.3$ (Figure~\ref{fig:atca_stack}), indicating an ESE was potentially in progress. On discovery of this ESE, we began high-cadence monitoring observations with ATCA, over a wider frequency range (2--11~GHz) than is possible in the survey mode. This monitoring revealed a double-horned light-curve (Figure \ref{fig:atca_ds}) between $t=0-100$ ($t=\rm{MJD}-56800$), with variations of almost a factor of three at some frequencies.  The shape of this light curve is strikingly similar to the light curves of some previous ESEs, although it does not have the short-lived spikes at 8~GHz observed for the archetypal ESE towards Q0954+658 \cite{Fiedler87ese}. These data reveal strong time variability, as is typical for ESEs. Some spectra are complex, containing occasionally two distinct peaks between 4 and 8~GHz (Figure \ref{fig:atca_stack}, 1~Jy$= 10^{-26} \unit{W~m^{-2}~Hz^{-1}}$).

PKS~1939$-$315 has been identified with a $V=20$~mag quasi-stellar object \cite{Mahony11} with unknown redshift. To check for temporary reddening from dust associated with the lens, we obtained $g'$, $r'$ and $i'$ band images with the Gemini South 8~m telescope.  The observations bracketed the second magnification event ($t=60$ to $t=100$) and revealed no variability in any band within a $3 \sigma$ limit of 0.5~mag \cite{SciSupplementary}. We also obtained 16 optical observations with the SMARTS 1.3~m telescope on a 3-day cadence during the second magnification event, in Johnston $V$ and Cousins $R$ and $I$ bands. These images also showed no evidence of variability within about 0.3~mag (3$\sigma$) \cite{SciSupplementary}.

To measure the geometry and angular scale of the lens, we obtained high-resolution, phase-referenced VLBI images with the Very Long Baseline Array (VLBA). We observed in 4 bands in the range 4--8~GHz with a $\sim 2$~day cadence, which sampled the second magnification event. We also obtained a 12~hr observation with the Australian Long Baseline Array during the first magnification event. Each VLBI image contained only one image of the background source. We detected significant, long-term astrometric shifts of the radio source during the ESE, which we conclude are the order of $\sim 1$~milliarcsecond \cite{SciSupplementary}.

The wide bandwidth of the ATCA radio data permits the study of the lens over a factor of $\sim10$ in wavelength. This corresponds to a factor of $\sim 10^2$ in the strength of the lens and provides strong constraints on lens models. Rather than restricting attention to a specific geometry and functional form for the electron column density ($N_e$) profile of the lens -- e.g. one-dimensional Gaussian \cite{Clegg98} -- we have developed a method for computing a 1D slice through the lens $N_e$ assuming only a geometry \cite{Tuntsov15, SciSupplementary}. The method relies on the fact that certain characteristic curves drawn through the dynamic spectrum each correspond to the same position in the lens plane. The lens properties can therefore be determined from the parameters of this family of curves.

We have applied this method assuming two different geometries (Fig.~S6): a highly anisotropic geometry (appropriate for modelling an edge-on sheet) and an axially-symmetric geometry (appropriate for modelling a spherical cloud or shell). We assume a distance of $1\,\mathrm{kpc}$  (e.g. \cite{Fiedler87ese}) and an effective transverse velocity of $50\,\mathrm{km}\,\mathrm{s}^{-1}$ (e.g. \cite{Rickett02})  for conversion to physical units.

\begin{figure}
\centering
\includegraphics[width=0.95\linewidth]{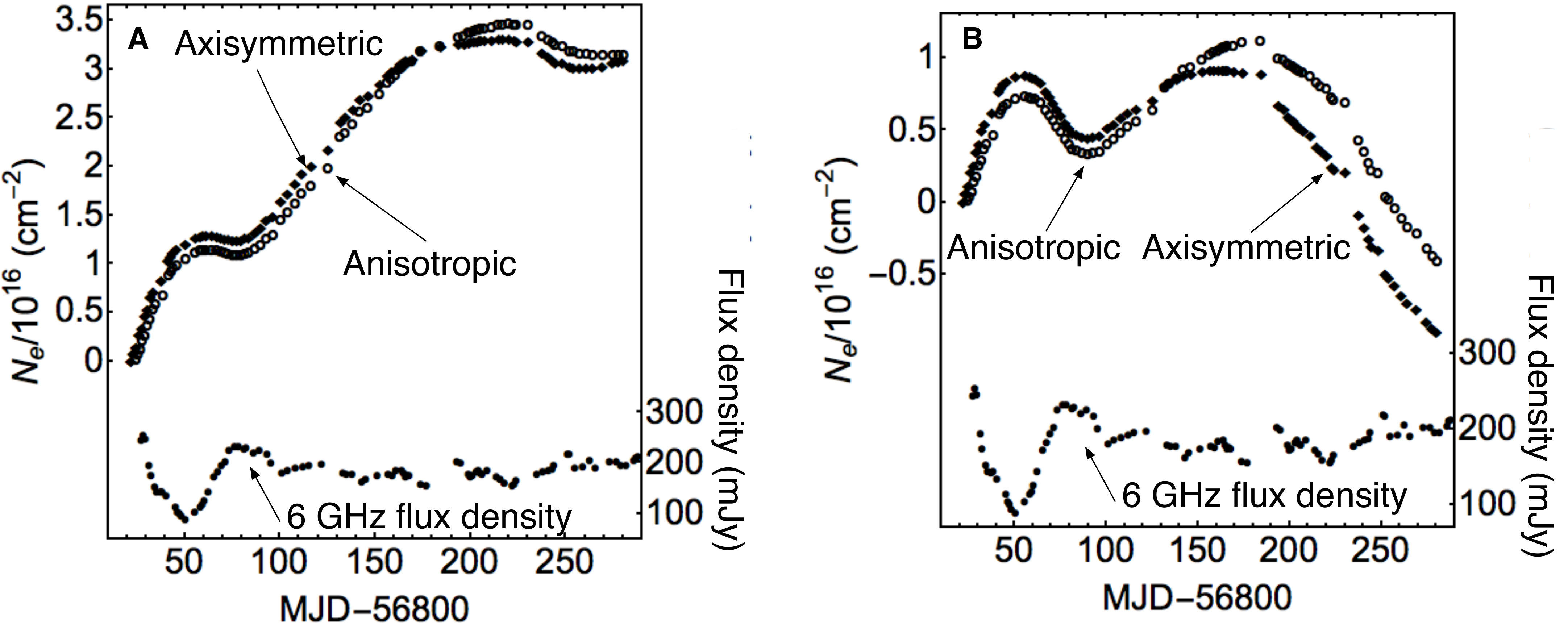}
\caption{\label{fig:ne_profile} {\bf The electron column density profile of the plasma lens.} These column densities were constructed by modelling the ATCA photometry (supplementary material), assuming axisymmetric and highly anisotropic geometries, a distance of 1~kpc and transverse velocity $V_{\text{eff}} = 50$~km/s. To facilitate comparison with the light curve, the density profiles are shown as a function of the position that each density measurement projects to at the source at 6~GHz. Profiles as a function of the lens coordinates are shown in Fig.~S8. Each point corresponds to a ``characteristic'' drawn through the dynamic spectrum (Fig.~S6), which has been identified with a particular position in the lens plane. The bottom plots are identical in both panels, and trace the 6~GHz light curve. Panel A: Modelled electron column density. Panel B: electron column density with a linear trend removed to emphasise electron column density excursions. The main ESE event, centred on $t=50$ corresponds to a local maximum in $N_e$, implying a divergent lens. The linear trends are $ 5.9 \times 10^{16}~\mathrm{cm^{-2}~yr^{-1}}$ and $5.2 \times 10^{16}~\mathrm{cm^{-2}~yr^{-1}}$ for the axisymmetric and extremely anisotropic cases respectively.  The gradient changes from $-4.3\times10^{16} \unit{cm^{-2}~yr^{-1}}$ (both geometries) to $2.4 \times 10^{17}$ ($1.9\times10^{17}$) $\unit{cm^{-2}~yr^{-1}}$  for the axisymmetric (anisotropic) geometry. This corresponds to a change in the deflection angle of 0.43 (0.35)~mas under the distance and velocity assumptions.  }
\end{figure}

The data do not distinguish between the two geometries. The resulting 1D slice through the electron column density of the lens looks very similar in both cases (Figure~\ref{fig:ne_profile}). The modelled electron column density increases approximately linearly with time, with the flux density variations being produced by the small-scale electron column density variations on top of a mean gradient. In both geometries the structure responsible for the deep flux minimum near $t=50$ is robustly modelled by a local maximum in the electron column, corresponding to a diverging lens. An over-focused, converging lens would simultaneously create a flux maximum and a focus at the higher frequencies within our observing band -- a possibility which is clearly excluded by our data.

For the axisymmetric model, the impact parameter is weakly constrained to be $0 \pm 10~\mathrm{days}$, corresponding to the lens centroid crossing the background source almost exactly. The most likely impact epoch is $t \approx -10$ and it is firmly earlier than $t \approx 0$. The peak in the $N_e$ profile is offset from this position, which implies either a shell-like, or aligned cylindrical morphology of the structure responsible for the ESE, rather than the density peaking at the symmetry centre. However, since the impact parameter is zero, the relative orientation of the effective velocity vector and the plasma density contours is constant, which is similar to the highly-anisotropic case.

We were unable to infer the geometry of the lens from the VLBI data, as multiple images of the background source were never observed, and the sizes of any time-dependent shifts were too small to be reliably detected. However, the long-lived shift of 1~mas is comparable to the shift inferred from change in the gradient of $N_e$ (Fig.~\ref{fig:ne_profile}) under our distance and velocity assumptions. Hereafter, we consider the implications of modelling the dynamic spectrum under both geometric assumptions. Under the anisotropic assumption, the modelling shows that the lens is diverging, rather than an over-focused converging one, which conclusively rules out the under-dense current-sheet model \cite{Pen12}. An over-dense sheet is consistent with the data. The conclusions under the axisymmetric assumption are less clear. A spherical cloud model would appear to be ruled out, due to the offset between the peak in the column density and the centre of symmetry. But, while a shell-like morphology implied in the axisymmetric model is qualitatively consistent with the cold-cloud prediction, where a cloud of self-gravitating neutral gas is surrounded by an ionised shell \cite{Henriksen95, Walker98, Walker07}, how the mean gradient fits into that picture is unclear.

Our best lens models exhibit column-density changes $\sim10^{16}\,{\rm cm^{-2}}$ over transverse scales $\sim10^{13}\,{\rm cm}$ (20 days at $50\,{\rm km\,s^{-1}}$ in Fig.~\ref{fig:atca_ds}, or see \cite{SciSupplementary} or \cite{Tuntsov15}), which yield gradients of $\sim10^{3}\,{\rm cm^{-3}}$. If the plasma structure responsible for the lensing is not highly elongated along the line-of-sight, that gradient should be comparable to the volume-density of electrons in the plasma. Its temperature should be  $T \gtrsim  3 \times 10^3$~K, for the material to be thermally ionised, which yields a plasma pressure of $2 n_e T \gtrsim 6 \times 10^6$~K~cm$^{-3}$, in accord with previous estimates of ESEs, and $\sim 2\times 10^3$ higher than the typical pressure in the diffuse interstellar medium \cite{Kulkarni88}.

In the case of PKS~1939$-$315 we have ruled out a converging lens, measured a VLBI astrometric shift, and determined the form of the column density profile. Ours and future surveys should generate a statistically significant sample of well-studied lenses, which will determine the overall covering fraction, optical depth and spatial distribution of ESE lenses. Some fraction of those ESEs should create multiple images of the background source resolvable with VLBI which will enable the geometry to be unambiguously determined. Such surveys may unveil a significant component of the Milky Way ISM.

\nocite{Fiedler87ese}
\nocite{Kulkarni88}
\nocite{Jenkins01,Jenkins11}
\nocite{Romani87,Rickett97}
\nocite{Hamidouche07}
\nocite{Pen12}
\nocite{Walker07}
\nocite{Henriksen95,Walker98,Walker07}
\nocite{Lazio01hi}
\nocite{Clegg96}
\nocite{Lazio00}
\nocite{Lazio01lc}
\nocite{Fiedler94}
\nocite{Cognard93,Cimo02,Maitia03,Savolainen08,Senkbeil08,Pushkarev13}
\nocite{Wilson11}
\nocite{Stix92}
\nocite{SciSupplementary}
\nocite{Valtaoja92}
\nocite{Fiedler87ese}
\nocite{Mahony11}
\nocite{SciSupplementary}
\nocite{SciSupplementary}
\nocite{SciSupplementary}
\nocite{Clegg98}
\nocite{Tuntsov15,SciSupplementary}
\nocite{Fiedler87ese}
\nocite{Rickett02}
\nocite{Pen12}
\nocite{Henriksen95,Walker98,Walker07}
\nocite{Kulkarni88}
\nocite{Bohlin78}
\nocite{Drake03}
\nocite{Fiedler87ese}
\nocite{Kulkarni88}
\nocite{Jenkins01,Jenkins11}
\nocite{Romani87,Rickett97}
\nocite{Hamidouche07}
\nocite{Pen12}
\nocite{Walker07}
\nocite{Henriksen95,Walker98,Walker07}
\nocite{Lazio01hi}
\nocite{Clegg96}
\nocite{Lazio00}
\nocite{Lazio01lc}
\nocite{Fiedler94}
\nocite{Cognard93,Cimo02,Maitia03,Savolainen08,Senkbeil08,Pushkarev13}
\nocite{Wilson11}
\nocite{Stix92}
\nocite{SciSupplementary}
\nocite{Valtaoja92}
\nocite{Fiedler87ese}
\nocite{Mahony11}
\nocite{SciSupplementary}
\nocite{SciSupplementary}
\nocite{SciSupplementary}
\nocite{Clegg98}
\nocite{Tuntsov15,SciSupplementary}
\nocite{Fiedler87ese}
\nocite{Rickett02}
\nocite{Pen12}
\nocite{Henriksen95,Walker98,Walker07}
\nocite{Kulkarni88}
\nocite{Bohlin78}
\nocite{Drake03}
\nocite{Fiedler87ese}
\nocite{Kulkarni88}
\nocite{Jenkins01,Jenkins11}
\nocite{Romani87,Rickett97}
\nocite{Hamidouche07}
\nocite{Pen12}
\nocite{Walker07}
\nocite{Henriksen95,Walker98,Walker07}
\nocite{Lazio01hi}
\nocite{Clegg96}
\nocite{Lazio00}
\nocite{Lazio01lc}
\nocite{Fiedler94}
\nocite{Cognard93,Cimo02,Maitia03,Savolainen08,Senkbeil08,Pushkarev13}
\nocite{Wilson11}
\nocite{Stix92}
\nocite{SciSupplementary}
\nocite{Valtaoja92}
\nocite{Fiedler87ese}
\nocite{Mahony11}
\nocite{SciSupplementary}
\nocite{SciSupplementary}
\nocite{SciSupplementary}
\nocite{Clegg98}
\nocite{Tuntsov15,SciSupplementary}
\nocite{Fiedler87ese}
\nocite{Rickett02}
\nocite{Pen12}
\nocite{Henriksen95,Walker98,Walker07}
\nocite{Kulkarni88}
\nocite{Bohlin78}
\nocite{Drake03}
\nocite{Fiedler87ese}
\nocite{Kulkarni88}
\nocite{Jenkins01,Jenkins11}
\nocite{Romani87,Rickett97}
\nocite{Hamidouche07}
\nocite{Pen12}
\nocite{Walker07}
\nocite{Henriksen95,Walker98,Walker07}
\nocite{Lazio01hi}
\nocite{Clegg96}
\nocite{Lazio00}
\nocite{Lazio01lc}
\nocite{Fiedler94}
\nocite{Cognard93,Cimo02,Maitia03,Savolainen08,Senkbeil08,Pushkarev13}
\nocite{Wilson11}
\nocite{Stix92}
\nocite{SciSupplementary}
\nocite{Valtaoja92}
\nocite{Fiedler87ese}
\nocite{Mahony11}
\nocite{SciSupplementary}
\nocite{SciSupplementary}
\nocite{SciSupplementary}
\nocite{Clegg98}
\nocite{Tuntsov15,SciSupplementary}
\nocite{Fiedler87ese}
\nocite{Rickett02}
\nocite{Pen12}
\nocite{Henriksen95,Walker98,Walker07}
\nocite{Kulkarni88}
\nocite{Bohlin78}
\nocite{Drake03}
\nocite{Fiedler87ese}
\nocite{Kulkarni88}
\nocite{Jenkins01,Jenkins11}
\nocite{Romani87,Rickett97}
\nocite{Hamidouche07}
\nocite{Pen12}
\nocite{Walker07}
\nocite{Henriksen95,Walker98,Walker07}
\nocite{Lazio01hi}
\nocite{Clegg96}
\nocite{Lazio00}
\nocite{Lazio01lc}
\nocite{Fiedler94}
\nocite{Cognard93,Cimo02,Maitia03,Savolainen08,Senkbeil08,Pushkarev13}
\nocite{Wilson11}
\nocite{Stix92}
\nocite{SciSupplementary}
\nocite{Valtaoja92}
\nocite{Fiedler87ese}
\nocite{Mahony11}
\nocite{SciSupplementary}
\nocite{SciSupplementary}
\nocite{SciSupplementary}
\nocite{Clegg98}
\nocite{Tuntsov15,SciSupplementary}
\nocite{Fiedler87ese}
\nocite{Rickett02}
\nocite{Pen12}
\nocite{Henriksen95,Walker98,Walker07}
\nocite{Kulkarni88}
\nocite{Bohlin78}
\nocite{Drake03}
\nocite{Fiedler87ese}
\nocite{Kulkarni88}
\nocite{Jenkins01,Jenkins11}
\nocite{Romani87,Rickett97}
\nocite{Hamidouche07}
\nocite{Pen12}
\nocite{Walker07}
\nocite{Henriksen95,Walker98,Walker07}
\nocite{Lazio01hi}
\nocite{Clegg96}
\nocite{Lazio00}
\nocite{Lazio01lc}
\nocite{Fiedler94}
\nocite{Cognard93,Cimo02,Maitia03,Savolainen08,Senkbeil08,Pushkarev13}
\nocite{Wilson11}
\nocite{Stix92}
\nocite{SciSupplementary}
\nocite{Valtaoja92}
\nocite{Fiedler87ese}
\nocite{Mahony11}
\nocite{SciSupplementary}
\nocite{SciSupplementary}
\nocite{SciSupplementary}
\nocite{Clegg98}
\nocite{Tuntsov15,SciSupplementary}
\nocite{Fiedler87ese}
\nocite{Rickett02}
\nocite{Pen12}
\nocite{Henriksen95,Walker98,Walker07}
\nocite{Kulkarni88}
\nocite{Bohlin78}
\nocite{Drake03}
\nocite{Fiedler87ese}
\nocite{Kulkarni88}
\nocite{Jenkins01,Jenkins11}
\nocite{Romani87,Rickett97}
\nocite{Hamidouche07}
\nocite{Pen12}
\nocite{Walker07}
\nocite{Henriksen95,Walker98,Walker07}
\nocite{Lazio01hi}
\nocite{Clegg96}
\nocite{Lazio00}
\nocite{Lazio01lc}
\nocite{Fiedler94}
\nocite{Cognard93,Cimo02,Maitia03,Savolainen08,Senkbeil08,Pushkarev13}
\nocite{Wilson11}
\nocite{Stix92}
\nocite{SciSupplementary}
\nocite{Valtaoja92}
\nocite{Fiedler87ese}
\nocite{Mahony11}
\nocite{SciSupplementary}
\nocite{SciSupplementary}
\nocite{SciSupplementary}
\nocite{Clegg98}
\nocite{Tuntsov15,SciSupplementary}
\nocite{Fiedler87ese}
\nocite{Rickett02}
\nocite{Pen12}
\nocite{Henriksen95,Walker98,Walker07}
\nocite{Kulkarni88}
\nocite{Bohlin78}
\nocite{Drake03}
\nocite{Wilson11}
\nocite{Murphy10}
\nocite{Wilson11}
\nocite{Sault95}
\nocite{Offringa10}
\nocite{Petrov11lba}
\nocite{Wrobel00}
\nocite{van-Moorsel96}
\nocite{Kettenis06}
\nocite{Pradel06}
\nocite{Schneider92}
\nocite{Tuntsov15}
\nocite{Deller11}
\nocite{Murphy10}
\nocite{Wilson11}
\nocite{Sault95}
\nocite{Offringa10}
\nocite{Petrov11lba}
\nocite{Wrobel00}
\nocite{van-Moorsel96}
\nocite{Kettenis06}
\nocite{Pradel06}
\nocite{Hook04}
\nocite{Schneider92}
\nocite{Tuntsov15}
\nocite{Deller11}
\nocite{Murphy10}
\nocite{Wilson11}
\nocite{Sault95}
\nocite{Offringa10}
\nocite{Petrov11lba}
\nocite{Wrobel00}
\nocite{van-Moorsel96}
\nocite{Kettenis06}
\nocite{Pradel06}
\nocite{Hook04}
\nocite{Schneider92}
\nocite{Tuntsov15}
\nocite{Deller11}
\nocite{Murphy10}
\nocite{Wilson11}
\nocite{Sault95}
\nocite{Offringa10}
\nocite{Petrov11lba}
\nocite{Wrobel00}
\nocite{van-Moorsel96}
\nocite{Kettenis06}
\nocite{Pradel06}
\nocite{Hook04}
\nocite{Schneider92}
\nocite{Tuntsov15}
\nocite{Deller11}
\nocite{Murphy10}
\nocite{Wilson11}
\nocite{Sault95}
\nocite{Offringa10}
\nocite{Petrov11lba}
\nocite{Wrobel00}
\nocite{van-Moorsel96}
\nocite{Kettenis06}
\nocite{Pradel06}
\nocite{Hook04}
\nocite{Schneider92}
\nocite{Tuntsov15}
\nocite{Deller11}
\nocite{Murphy10}
\nocite{Wilson11}
\nocite{Sault95}
\nocite{Offringa10}
\nocite{Petrov11lba}
\nocite{Wrobel00}
\nocite{van-Moorsel96}
\nocite{Kettenis06}
\nocite{Pradel06}
\nocite{Hook04}
\nocite{Schneider92}
\nocite{Tuntsov15}
\nocite{Deller11}
\nocite{Murphy10}
\nocite{Wilson11}
\nocite{Sault95}
\nocite{Offringa10}
\nocite{Petrov11lba}
\nocite{Wrobel00}
\nocite{van-Moorsel96}
\nocite{Kettenis06}
\nocite{Pradel06}
\nocite{Hook04}
\nocite{Schneider92}
\nocite{Tuntsov15}
\nocite{Deller11}

\bibliography{Master.bib}

\bibliographystyle{Science}

\subsection*{Acknowledgements}

We acknowledge valuable discussions with Ron Ekers in the preparation of this paper.

We thank the ATNF scheduler, Phil Edwards, for making the regular ATCA and LBA observations possible, and the staff at Narrabri for facilitating access to the ATCA.

Full acknowledgements are available in the supplementary materials.

All data described in this paper are attached to the Supporting Online Material.

The authors report no conflicts of interest.

\subsection*{Supplementary materials}
Materials and Methods
Figs. S1--S9.
Tables. S1
References: 26-42

\clearpage

\end{document}